\documentclass[aps,prc,twocolumn,showpacs,amsmath,amssymb,nofootinbib]{revtex4-1}
\usepackage{blindtext}
\usepackage{graphicx}
\usepackage{color}
\usepackage{times}
\usepackage{bm}
\usepackage{multirow}
\usepackage{float}
\usepackage{url}
\usepackage{natbib}
\usepackage{tensor}
\usepackage{bbm}
\usepackage[colorlinks=true,citecolor=blue,urlcolor=blue,linkcolor=red]{hyperref}
\usepackage[normalem]{ulem}
\usepackage{verbatim}
\usepackage{lipsum} 
\usepackage{tikz,xcolor,hyperref}
\usepackage{ifxetex,ifluatex}
\usepackage{booktabs}

\newcommand{\physrep}{\textit{Phys.~Rep.}}
\newcommand{\nphysa}{\textit{Nucl.~Phys.~A}}

\definecolor{lime}{HTML}{A6CE39}
\DeclareRobustCommand{\orcidicon}{%
	\begin{tikzpicture}
	\draw[lime, fill=lime] (0,0) 
	circle [radius=0.17] 
	node[white] {{\fontfamily{qag}\selectfont \tiny ID}};
	\draw[white, fill=white] (-0.0625,0.095) 
	circle [radius=0.008];
	\end{tikzpicture}
	\hspace{-2mm}
}
\foreach \x in {A, ..., Z}{
	\expandafter\xdef\csname orcid\x\endcsname{\noexpand\href{https://orcid.org/\csname orcidauthor\x\endcsname}{\noexpand\orcidicon}}
}

\begin{document}

\title{Bayesian Analyses of Proton Multiple Flow Components in Intermediate Heavy Ion Collisions with Momentum-Dependent Interactions}
\author{Shuochong Han$^{1}$}
\author{Ang Li{\orcidA{}}$^{1}$}
\affiliation{\it $^{1}$Department of Astronomy, Xiamen University, Xiamen 361005, China; \textit{liang@xmu.edu.cn}
}

\begin{abstract} 
We perform a comprehensive Bayesian analyses of Au + Au collision data at 1.23 GeV/nucleon using an isospin-dependent Boltzmann-Uehling-Uhlenbeck transport model that incorporates a momentum-dependent mean field and medium-modified baryon-baryon cross sections. The model parameters are calibrated to empirical properties of nuclear matter at saturation density, with particular attention to variations in the incompressibility $K_0$. 
Within a Bayesian statistical framework and using a Gaussian Process emulator, we simultaneously extract constraints on the incompressibility $K_0$ and the in-medium baryon-baryon scattering modification factor $X$ by systematically comparing model predictions with HADES measurements of proton collective flow, including the slopes ($F_1$ and $F_3$) of directed and triangular flow, as well as elliptic ($v_2$) and quadrupole ($v_4$) flow observables.
We find that the extracted incompressibility favors relatively small values, indicating a soft nuclear equation of state, while the inferred average $X$ values fall at $0.9$--$ 1.0$, suggesting mild suppression of baryon-baryon cross sections in the medium. 
Furthermore, we demonstrate that transport models employing momentum-independent mean fields require stiffer equations of state and stronger in-medium corrections to reproduce the same observables. These results highlight the critical role of momentum dependence in the mean field and its interplay with in-medium scattering in constraining the properties of dense nuclear matter from heavy-ion collisions.

\end{abstract}

\maketitle 

\section{Introduction}

The single-particle mean field and the in-medium baryon-baryon scattering cross sections are two fundamental inputs in transport models used to simulate heavy-ion collisions (HICs). 
These two inputs are closely linked to the nuclear equation of state (EoS) and are essential for accurately modeling the dynamical evolution of nuclear matter under extreme conditions.
Theoretically both the single-particle mean field and the in-medium nucleon-nucleon ($NN$) cross sections can be self-consistently determined from microscopic many-body approaches, such as the Brueckner-Hartree-Fock (BHF)~\cite{1996PhRvC..53.1478G,1997PhRvC..55.3006S,2002EPJA...14..469Z,2005PhRvC..72a4005Z,2007PhRvC..76e4001Z}, Dirac-BHF (DBHF)~\cite{1993PhRvC..48.1702L,1994PhRvC..49..566L,2025PhRvC.112e4322W}, and variational approaches~\cite{1992PhRvC..45..791P}, which are based on bare $NN$ interactions. 
The single-particle mean field is directly related to the EoS through the corresponding energy density functional~\cite{1988PhR...160..189B,1991PhR...202..233A,2025PhRvC.111e4605W}.
In intermediate-energy HICs, the development of collective flow is jointly influenced by the stiffness of the nuclear EoS and the in-medium baryon-baryon cross sections. 
However, the implementation of both the microscopically calculated mean field and in-medium baryon-baryon cross sections in transport simulations remains a longstanding goal yet to be fully realized.

Experimentally, various components of the nucleon collective flow extracted from HIC experiments provide crucial insights into both the mean field and the in-medium $NN$ cross sections~\cite{1993PhRvL..71.1986W,2009PrPNP..62..419C,2018PhLB..778..207W,2022PhLB..82837019L,2022PhLB..82937134G,2023NuPhA103922726L}. 
A common strategy in current transport studies is to vary either the EoS or the in-medium cross section independently and compare model predictions with experimental data to extract information about nuclear matter properties.
However, since these two inputs are tightly coupled in their impact on observables, treating them independently introduces ambiguities and often prevents a reliable, simultaneous constraint of both, highlighting the need for more integrated approaches in the interpretation of HIC data~\cite{2005PhRvC..72f4611L,2007PhRvC..75c4615Z,2011PhRvC..83d4617L}.

The stiffness of the EoS for cold, symmetric nuclear matter is characterized by the incompressibility $K_0$ at saturation density $\rho_0$. 
Accurately determining the value of $K_0$ is of great significance for understanding a wide range of phenomena, including nuclear structure, nuclear reactions, and the properties of neutron stars. 
Over the past few decades, significant progress has been made in constraining $K_0$ through various approaches, such as analyses of the giant monopole resonance (GMR)~\cite{1980PhR....64..171B,2010JPhG...37f4038P,2014PhRvC..89d4316S} in finite nuclei, collective flow and particle yields~\cite{2016NuPhA.945..112L,2018PhLB..778..207W,2024PhRvC.109e4619L} in HICs, as well as neutron star mass and radius measurements~\cite{2018PhRvC..97c5805Z,2024PhRvD.110j3040L}. 
In HICs, the incompressibility $K_0$ determines the maximum density achieved in the reaction zone, thereby influencing the collective flow of final-state particles.  
Meanwhile, the cross section governs the frequency of collisions during the reaction and strongly affects the final-state particle distributions and yields.
The in-medium $NN$ cross section is modified by the nuclear environment through two main mechanisms~\cite{1997PhRvC..55.3006S,2022PhRvC.106f4332H,2025arXiv250723476H}: changes in the scattering amplitude and alterations in the density of states. 
In medium $NN$ cross sections calculated with microscopic many-body approaches such as BHF and DBHF, the medium effect of the scattering amplitude arises from the mean field and the Pauli blocking of the intermediate states, while the density of states is affected via modifications to the nucleon effective mass.
BHF studies indicate that the medium effect on the scattering amplitude tends to enhance the cross section, while the correction from the effective mass suppresses it~\cite{1997PhRvC..55.3006S,2022PhRvC.106f4332H,Sun2025Astronomical,2025arXiv250723476H}. 
Importantly, the suppression due to the effective mass is significantly stronger than the enhancement from the scattering amplitude~\cite{2022PhRvC.106f4332H}. 
As a result, in many transport models, it is reasonable to treat the scattering amplitude as identical to that in free space and only consider the medium correction through the effective mass. 

In recent years, Bayesian inference has gained prominence as a powerful and statistically rigorous tool in nuclear physics, offering a systematic framework to quantify uncertainties and disentangle the effects of correlated physical parameters~\cite{2021ApJ...913...27L,2023NuPhA103922726L,2023PhRvC.107e5803Z,2023ApJ...943..163Z,2023ApJ...942...55S,2024arXiv240607051W,2024PhRvD.109l3005M,2025PhRvC.111e4602L}. 
Previous approaches typically involved varying one model parameter at a time while keeping others fixed. 
Such methods often lead to biased or ambiguous interpretations, particularly when the influences of the nuclear matter EoS and in-medium cross sections on observables are intrinsically coupled.

In this study, we perform a joint Bayesian analysis of the EoS and in-medium cross sections by using a Gaussian Process (GP) emulator to simulate the collective flows predicted by the isospin-dependent Boltzmann-Uehling-Uhlenbeck (IBUU) transport model.
The reaction system considered is Au+Au at a beam energy of 1.23 GeV per nucleon. 
For experimental constraints, we adopt the proton collective flow data measured by the HADES Collaboration~\cite{2017NuPhA.967..812K,2020PhRvL.125z2301A,2023EPJA...59...80A}.
In the IBUU model~\cite{2004NuPhA.735..563L}, we implement a momentum-dependent mean-field potential for nucleons. 
The collective flow observables employed in our analysis include the slope of directed flow ($F_1$), the elliptic flow ($v_2$), the slope of triangular flow ($F_3$), and the quadrupole flow ($v_4$).
This Bayesian framework enables a simultaneous extraction of both the in-medium correction factor for cross sections and the incompressibility $K_0$ of symmetric nuclear matter. 

The structure of this paper is organized as follows: 
In Sec.~\ref{Sec:model}, we briefly describe the key components of the IBUU transport model and the setup details for simulating Au+Au collisions. 
Sec.~\ref{Sec:Bay} outlines the essential aspects of the Bayesian inference framework adopted in this study. 
Sec.~\ref{Sec:Result} is divided into two parts: the first presents and analyzes the collective flow results predicted by the IBUU transport model; the second evaluates the performance of the GP emulator by comparing its outputs with IBUU simulations. This section also provides the posterior probability distribution functions (PDFs) for the incompressibility $K_0$ and the in-medium cross-section correction factor $X$, along with the corresponding Bayesian inference parameters under different scenarios. 
Sec.~\ref{Sec:Summary} provides a summary of the main findings and conclusions of this work.

\section{IBUU Transport Model}\label{Sec:model}

\begin{table*}
\squeezetable
\caption{The values of $A_{l0}$, $A_{u0}$, $B$, $\sigma$, $C_{l0}$, $C_{u0}$, and $\Lambda_0$ for different $K_0$ values in the isospin- and momentum-dependent single-particle potential energy (Eq.~\ref{1}).} \label{parameters}
\setlength{\tabcolsep}{1.8mm}
\renewcommand\arraystretch{1.5}
\begin{ruledtabular}
\begin{tabular*}{\hsize}{@{}@{\extracolsep{\fill}}lcccccccc@{}}
&$K_0~\mathrm{(MeV)}$&$A_{l0}=A_{u0}~\mathrm{(MeV)}$&$B~\mathrm{(MeV)}$&$C_{l0}~\mathrm{(MeV)}$&$C_{u0}~\mathrm{(MeV)}$&$\Lambda_0$&$\sigma$\\
\hline
&150&181.44&-106.44&-60.54&-99.74&2.42&0.73\\
\hline
&170&301.78&-226.78&-60.54&-99.74&2.42&0.87\\
\hline
&190&11030.67&-10955.67&-60.54&-99.74&2.42&0.10\\
\hline
&210&-196.34&271.34&-60.54&-99.74&2.42&1.13\\
\hline
&230&-67.52&142.52&-60.54&-99.74&2.42&1.26\\
\hline
&250&-25.24&100.24&-60.54&-99.74&2.42&1.40\\
\hline
&270&-4.22&79.22&-60.54&-99.74&2.42&1.53\\
\hline
&290&8.36&66.64&-60.54&-99.74&2.42&1.67\\
\end{tabular*}
\end{ruledtabular}
\end{table*}

The present IBUU transport model has its origin from the IBUU04 model~\cite{2004NuPhA.735..563L}. 
The Boltzmann-Uehling-Uhlenbeck (BUU) transport model describes the time evolution of the one-body phase-space distribution function $f_{\tau}(\vec{r},\vec{p},t)$, which is expressed as:
\begin{equation}
    \frac{\partial f}{\partial t}+\nabla_{\vec{p}}E\cdot\nabla_{\vec{r}}f-\nabla_{\vec{r}}E\cdot\nabla_{\vec{p}}f=I_c,\label{Eq:BUU}
\end{equation}
where the left-hand side denotes the times evolution of the particle phase-space distribution function due to its transport and mean field, while the right-hand side, $I_c$, is the collision integral accounting for changes in $f$ due to elastic and inelastic two-body collisions. 
The total energy $E$ of a particle consists of its kinetic energy $E_{\rm{kin}}$ and its mean-field potential energy $U$: $E=E_{\rm{kin}}+U$.

The isospin- and momentum-dependent single-particle potential energy is expressed as~\cite{1987PhRvL..58.1926A,1987PhRvC..35.1666G,1988PhRvC..38.2101W,1988PhRvC..38.2101W,1993PhRvL..70.2062P,2003PhRvC..67c4611D,2004NuPhA.735..563L,2015PhRvC..91a4611X,2017PhRvC..95c4324K,2017PhRvC..96d4605Y}:
\begin{align}
    U_{\tau}(\rho,\beta,\vec{p})&=A_u\frac{\rho_{-\tau}}{\rho_0}+A_l\frac{\rho_{\tau}}{\rho_0}+B\Big(\frac{\rho}{\rho_0}\Big)^{\sigma}(1-x\beta^2)\nonumber \\
    &-8\tau x\frac{B}{\sigma+1}\frac{\rho^{\sigma-1}}{\rho_0^{\sigma}}\beta\rho_{-\tau}\nonumber \\
    &+\frac{2C_l}{\rho_0}\int d^3p'\frac{f_{\tau}(\vec{r},\vec{p}')}{1+(\vec{p}-\vec{p}')^2/\Lambda^2}\nonumber \\
    &+\frac{2C_u}{\rho_0}\int d^3p'\frac{f_{-\tau}(\vec{r},\vec{p}')}{1+(\vec{p}-\vec{p}')^2/\Lambda^2}\label{1},
\end{align}
where $\rho_0$ denotes the saturation density, $\tau=\pm\frac{1}{2}$ is for the neutron/proton potential, $\beta=(\rho_n-\rho_p)/(\rho_n+\rho_p)$ is the isospin asymmetry, and $\rho_n$ and $\rho_p$ is the neutron and proton densities, respectively. 
In the isospin and momentum-dependent interaction (ImMDI)~\cite{2015PhRvC..91a4611X,2017PhRvC..95c4324K}, the isovector parameters $x$, $y$, and $z$ are introduced, and $A_l$, $A_u$, $C_l$, and $C_u$ can then be expressed as:
\begin{align}
    A_l&=A_{l0}+y+x\frac{2B}{\sigma+1},\\
    A_u&=A_{u0}-y-x\frac{2B}{\sigma+1},\\
    C_l&=C_{l0}-2(y-2z)\frac{p^2_{f_0}}{\Lambda^2 ln[(4p^2_{f_0}+\Lambda^2)/\Lambda^2]}, \\
    C_u&=C_{u0}+2(y-2z)\frac{p^2_{f_0}}{\Lambda^2 ln[(4p^2_{f_0}+\Lambda^2)/\Lambda^2]}, 
\end{align}
where $p_{f_0}=\hbar(3\pi^2\rho_0/2)^{1/3}$ is the nucleon Fermi momentum in symmetric nuclear matter at the saturation density. 
Here we take $\Lambda=\Lambda_0 p_{f_0}$. 
The seven parameters embedded in above expressions, i.e., $A_{l0}=A_{u0}$, $B$, $\sigma$, $C_{l0}$, $C_{u0}$ and $\Lambda_0$, are determined by fitting six experimental and/or empirical constraints on the properties of nuclear matter at $\rho_0=0.16~\mathrm{fm^{-3}}$. 
Specifically, for $x=y=z=0$, we choose the following empirical values, i.e., the saturation density $\rho_0=0.16~\mathrm{fm^{-3}}$, the binding energy $E_0(\rho_0)=-16~\mathrm{MeV}$, the isoscalar effective mass $m^*=0.7m$, the symmetry energy $E_{\rm{sym}}(\rho_0)=32.5~\mathrm{MeV}$, and the single-particle potential at infinite momentum: $U^{\infty}_0=75~\mathrm{MeV}$, evaluated in nuclear matter at saturation density, as detailed in the Appendix \ref{appendixA}. 
The corresponding values of the parameters for different incompressibility $K_0$ are listed in Table \ref{parameters}. 
From these results, we observe that the parameters $C_{l0}$, $C_{u0}$, and $\Lambda_0$ are 
nearly independent of the value of $K_0$, and can thus be treated as constants in our analysis. 
In the transport simulations of Au + Au HICs, the isovector-vector parameters used are chosen as $x = 1.0,~ y = -100\ \mathrm{MeV},~ z = -2.5\ \mathrm{MeV}$~\cite{2015PhRvC..91d7601K,2019PhRvC.100b4618X,2020PhRvC.102b4306X}.
Based on Eqs. (\ref{Eq:dEsym_drho}) and (\ref{Eq:L0}), we calculated the symmetry-energy slope $L_0$ for each parameter set, and found that all $L_0$ values are nearly identical, approximately 6.13 MeV.

The formula for the compressibility $K_0$ of cold matter at $\rho=\rho_0$ is
\begin{align}
    K_0&=9\frac{d P}{d\rho}\Big|_{\rho=\rho_0}\nonumber\\
    &=\Big[18\rho\frac{dE_0(\rho)}{d\rho}+9\rho^2\Big(\frac{d^2E_0(\rho)}{d\rho^2} \Big)\Big]_{\rho=\rho_0} \nonumber\\
    &=\frac{3p_{f_0}^2}{m}+9A_{l0}+9B\sigma\nonumber\\
    &+\frac{27\Lambda_0^2(C_{l0}+C_{u0})}{8}\Big[\frac{4}{3}-\frac{\Lambda_0^2}{3}ln\frac{4+\Lambda_0^2}{\Lambda_0^2} \Big]\nonumber\\
    &=\frac{3p_{f_0}^2}{m}+9A_{l0}+9B\sigma-1005.21 ~\mathrm{(MeV)}，
\end{align}
where $P$ denotes pressure.
Therefore, the correction introduced by the momentum-dependent part is approximately -1005 MeV. 
This indicates a stronger momentum dependence compared to the -838.76 MeV correction derived from Eq.~(13) in Ref.~\cite{1987PhRvC..35.1666G}.
Considering that the first derivative of the EoS of symmetric nuclear matter with respect to density vanishes at the saturation density $\rho_0$, i.e. $\frac{dE_0(\rho)}{d\rho}\Big|_{\rho=\rho_0}=0$, the incompressibility $K_0$ can be expressed as
\begin{align}
    K_0&=9\rho_0^2\Big(\frac{d^2E_0(\rho)}{d\rho^2} \Big)_{\rho=\rho_0}\nonumber\\
    &=-\frac{3p_{f_0}^2}{5m}+\frac{9B(\sigma-1)\sigma}{\sigma+1}+85.03~\mathrm{(MeV)}.
\end{align}
Finally, the momentum dependence of the mean field contributes a correction of 85.03 MeV to the incompressibility $K_0$.
Consequently, we obtained the following relationship for the parameters $A_{l0}$, $B$ and $\sigma$ with respect to the incompressibility $K_0$:
\begin{align}
    A_{l0}&=75-16.6306\frac{K_0+108.9012}{K_0-190.4496} ~\mathrm{(MeV)}, \\
    B&=16.6306\frac{K_0+108.9012}{K_0-190.4496} ~\mathrm{(MeV)}, \\
    \sigma&=\frac{K_0-40.7742}{149.6754} .
\end{align}

For comparison, we also perform an analysis using a momentum-independent mean field. The adopted expression for the momentum-independent potential is
\begin{align}
    U_q(\rho,\beta)=a(\frac{\rho}{\rho_0})+b(\frac{\rho}{\rho_0})^{\alpha}+U_{asy}^q+U_C^q.
\end{align}
Here, $U_{asy}^q$ is the symmetric potential for baryon $q$, $U_C^q$ is the Coulomb interaction between charged particles. The parameters $a,~ b$ and $\alpha$ are determined by the properties of symmetric nuclear matter at the saturation point. 
Their relationship with the incompressibility $K_0$ can be expressed as~\cite{1988PhR...160..189B,1998IJMPE...7..147L}
\begin{align}
    a&=-29.81-46.90\frac{K_0+44.73}{K_0-166.32}\mathrm{(MeV)},\nonumber\\
    b&=23.45\frac{K_0+255.78}{K_0-166.32}\mathrm{(MeV)},\nonumber\\
    \alpha&=\frac{K_0+44.73}{211.05}.
\end{align}

The in-medium baryon-baryon scattering cross section is scaled down by the parameter $X$~\cite{2023NuPhA103922726L,2025PhRvC.111e4602L}, 
\begin{equation}
X=\sigma^{\rm{medium}}_{\rm{BB_{elastic,inelastic}}}/\sigma^{\rm{free}}_{\rm{BB_{elasic,inelasic}}}, 
\end{equation}
where $\sigma^{\rm{free}}_{\rm{BB_{elasic,inelasic}}}$ is the total baryon-baryon scattering cross section in free space. 
In free space, the $NN$ inelastic isospin-decomposed cross sections are given by:
\begin{align}
    \sigma^{pp\rightarrow n\Delta^{++}}&=\sigma^{nn\rightarrow p\Delta^-}=\sigma_{10}+\frac{1}{2}\sigma_{11},\nonumber\\
    \sigma^{pp\rightarrow p\Delta^{+}}&=\sigma^{nn\rightarrow n\Delta^0}=\frac{3}{2}\sigma_{11},\nonumber\\
    \sigma^{np\rightarrow p\Delta^{0}}&=\sigma^{np\rightarrow n\Delta^+}=\frac{1}{2}\sigma_{11}+\frac{1}{4}\sigma_{10},    
\end{align}
where $\sigma_{II'}$ follows the parameterization given in Ref. ~\cite{1982PhRvC..25.1979V}. Here, $I$ and $I'$ denote the isospin of the initial and final collision particles, respectively.
The cross section for the two-body free inverse reaction $N\Delta\rightarrow NN$ is calculated by the modified detailed balance.

We then simulated the Au + Au reaction at $E_{\mathrm{beam}} = 1.23$ GeV by employing the IBUU transport model for two centrality $(10$--$20)\%$ (the impact parameter $b=4.7$--$6.6~\mathrm{fm}$) and $(20$--$30)\%$ (the impact parameter $b=6.6$--$8.1~\mathrm{fm}$). 
The probability density $P(b) \propto b$ of the impact parameter $b$.
For each set of parameters $X$ and $K_0$, we have set up 250 testparticles/nucleon and randomly generated 200 collision parameters.

The four observables we employed in this work are the slope $F_1=dv_1/dy|_{y=0}$ of the direct flow $v_1$, the elliptic flow $v_2$, the slope $F_3=dv_3/dy|_{y=0}$ of the triangular flow $v_3$ and quadrangular flow $v_4$ for the proton. 
The anisotropic flow $v_n$ of particles are the Fourier coefficients in the decomposition of their transverse momentum spectra in the azimuthal angle $\phi$ with respect to the reaction plane~\cite{1998PhRvC..58.1671P}, i.e., 
\begin{align}
    E\frac{d^3N}{dp^3}=\frac{1}{2\pi}\frac{d^2N}{p_tdp_tdy}\Big[1+\sum_{n=1}^{\infty}2v_n(p_t,y)\mathrm{cos}(n\phi) \Big].
\end{align}
In this work, the reaction plane is in the $x-o-z$ plane. 
The flow coefficients $v_n$ of order $n$ are defined as
\begin{align}
    v_n=\langle \mathrm{cos}(n\phi)\rangle. 
\end{align}
Here, $\langle ...\rangle$ denotes the average over all selected particles and events in a given sample. 
Further, the anisotropic flows $v_n$ as functions of the rapidity $y$ and the transverse momentum $p_t$ can be expressed as~\cite{2004PhRvC..69c1901C,2005PhLB..605...95C,2025PhRvC.111e4605W} 
\begin{align}
    v_1(y,p_t)&=\frac{1}{n}\sum_{i=1}^{n}\frac{p_{x_i}}{p_{t_i}},\\
    v_2(y,p_t)&=\frac{1}{n}\sum_{i=1}^{n}\frac{p_{x_i}^2-p_{y_i}^2}{p_{t_i}^2},\\
    v_3(y,p_t)&=\frac{1}{n}\sum_{i=1}\frac{p_{x_i}^3-3p_{x_i}p_{y_i}^2}{p_{t_i}^3},\\
    v_4(y,p_t)&=\frac{1}{n}\sum_{i=1}\frac{p_{x_i}^4-6p_{x_i}^2p_{y_i}^2+p_{y_i}^4}{p_{t_i}^4},
\end{align} 
where $y=y_{\rm cm}/y_{\rm mid}$ in the center of mass (cm) frame of the two colliding nuclei, and $y_{\rm mid}=0.74$. 

\section{Bayesian Inference}\label{Sec:Bay}

Within the framework of Bayesian parameter estimation, the probability distribution of the model parameters $\theta$ conditioned on the observed data $D$ (referred to as the posterior distribution) is given by: 
\begin{equation}
    P(\theta|D)=\frac{P(D|\theta)P(\theta)}{\int P(D|\theta)P(\theta)d\theta},
\end{equation}
where $P(\theta)$ is the prior distribution. 
The likelihood of the full data set $D$ given parameters $\theta$ is obtained by multiplying the likelihoods of the individual data points: 
\begin{equation}
    P(D|\theta)=\prod_i P(d_i|\theta). 
\end{equation}
The likelihood function at individual points is given by a Gaussian distribution
\begin{equation}
   P(d_i|\theta) = \frac{1}{\sqrt{2\pi}\sigma_i}\mathrm{exp}\Big[ -\frac{1}{2}\Big( \frac{d_i-\xi_i(\theta)}{\sigma_i} \Big)^2 \Big],
\end{equation}
where $d_i$ denotes HADES experimental data and $\xi_i(\theta)$ represents simulated data. The error term is $\sigma_i=\sqrt{\sigma^2_{\mathrm{theory}}+\sigma^2_{\mathrm{experiment}}+\sigma^2_{\mathrm{GP}}}$, where $\sigma_{\mathrm{theory}}$ denotes the IBUU model statistical error, $\sigma_{\mathrm{experiment}}$ represents the HADES experimental error, and $\sigma_{\mathrm{GP}}$ is the error from the Gaussian process. 
Because the statistical uncertainties associated with different ($X$ and $K_0$) points differ only slightly, we use their averaged values as part of the model uncertainty when constructing the likelihood function.
The averaged IBUU model statistical errors ($\sigma_{\mathrm{theory}}$) for $F_1$, $v_2$, $F_3$, and $v_4$ are approximately 0.011, 0.0022, 0.0072, and 0.0022, respectively. 
Correspondingly, the averaged GP errors ($\sigma_{\mathrm{GP}}$) for these observables are approximately 0.008, 0.003, 0.0086, and 0.003.

In the present work, we investigate the collective flow of protons in Au + Au collisions at two centralities, using experimental data from HADES~\cite{2023EPJA...59...80A,2017NuPhA.967..812K,2020PhRvL.125z2301A}. 
Specifically, we use the slopes $F_1$ and $F_3$ of the directed flow $v_1$ and triangular flow $v_3$ at mid-rapidity ($y\equiv y_{\rm{cm}}/y_{\rm{mid}}=0$), as well as the elliptic $v_2$ and quadrupole flow $v_4$ values at a transverse momentum of $0.6~\mathrm{GeV/c}\leq p_t \leq 0.9~\mathrm{GeV/c}$ .
Details of the experimental data are collected in Table~\ref{Tab:exp data}.

\begin{table}
\squeezetable
\centering
\caption{Experimental values for the collective flow of protons from HADES~\cite{2023EPJA...59...80A}.}
\label{Tab:exp data}
\setlength{\tabcolsep}{1.8mm}
\renewcommand\arraystretch{1.5}
\begin{ruledtabular}
\begin{tabular}{@{}lcc@{}}
\textbf{Flow} & $10$--$20\%(\times 10^{-2})$ & $20$--$30\%(\times 10^{-2})$\\
\hline
$F_1$ & $54.49 \pm 3.96$ & $57.01 \pm 3.51$ \\
$v_2$ & $-8.03 \pm 0.55$ & $-11.20 \pm 0.41$ \\
$F_3$ & $-4.87 \pm 0.35$ & $-7.01 \pm 0.36$ \\
$v_4$ & $0.41 \pm 0.12$ & $0.70 \pm 0.11$ \\
\end{tabular}
\end{ruledtabular}
\end{table}

As is well known, the incompressibility $K_0$ of symmetric nuclear matter still carries considerable uncertainty. 
Based on constraints from HICs and giant monopole resonances, its widely accepted range lies approximately between 180 and 300~MeV~\cite{2006EPJA...30...23S,2014PhRvC..89d4316S,2018PhRvC..98e4316B,2018PhLB..778..207W,2018PrPNP.101...55G}. 
Therefore, in this study, we adopt a relatively broad prior distribution for $K_0$, ranging from 100 to 380 MeV.
The parameter $X$ characterizes the medium modification of the in-medium baryon-baryon scattering cross section relative to that in vacuum. 
Current studies based on microscopic many-body approaches~\cite{1997PhRvC..55.3006S,2007PhRvC..76e4001Z,2022PhRvC.106f4332H,2025PhRvC.112e4322W} suggest that medium effects tend to suppress the $NN$ elastic scattering cross section, i.e., the ratio $\sigma_{\text{medium}} / \sigma_{\text{free}}$ is expected to be less than 1.0.
In our previous work, based on microscopic BHF calculations~\cite{2022PhRvC.106f4332H,2025arXiv250723476H}, we found that the modification factor of the in-medium elastic $NN$ cross section increases with laboratory energy, indicating a weakening of medium suppression at higher energies. 
Nevertheless, the presence of inelastic $NN$ processes introduces further complexity to particle collisions in the medium, increasing the uncertainty in modeling the transport dynamics.
For inelastic scattering, there is currently no consensus. 
Taking all the above into account, we adopt a relatively broad prior distribution for the medium modification factor $X$ over the range from 0.5 to 2.0. 
To summarize, we use uniform priors $U[100,380]$ MeV for $K_0$ and $U[0.5,2.0]$ for $X$ in our analysis. 
To assess the specific impact of momentum dependence, here we performed a parallel analysis with a momentum-independent mean field while keeping all other inputs identical. The resulting posteriors of $K_0$ and $X$ are shown below for the cases with and without the MDI (momentum-dependent interaction). The contrast is striking: without the MDI, reproducing the HADES flow data requires substantially larger values of $X \sim 1.3$--$1.6$ and $K_0 > 240$ MeV, whereas with the MDI the favored range is $X \sim 0.9$--$1.0$ and $K_0 < 210$ MeV. 

\section{Results and discussion}\label{Sec:Result}

This section is divided into two parts. 
In the first part, we investigate how the in-medium cross-section modification factor $X$ and the incompressibility $K_0$ influence the results of IBUU transport model simulations. 
In the second part, we present the Bayesian posterior distributions obtained under constraints from experimental data.

\subsection{IBUU transport model simulation}

\begin{figure*}
    \centering    
    \includegraphics[width=0.9\textwidth]{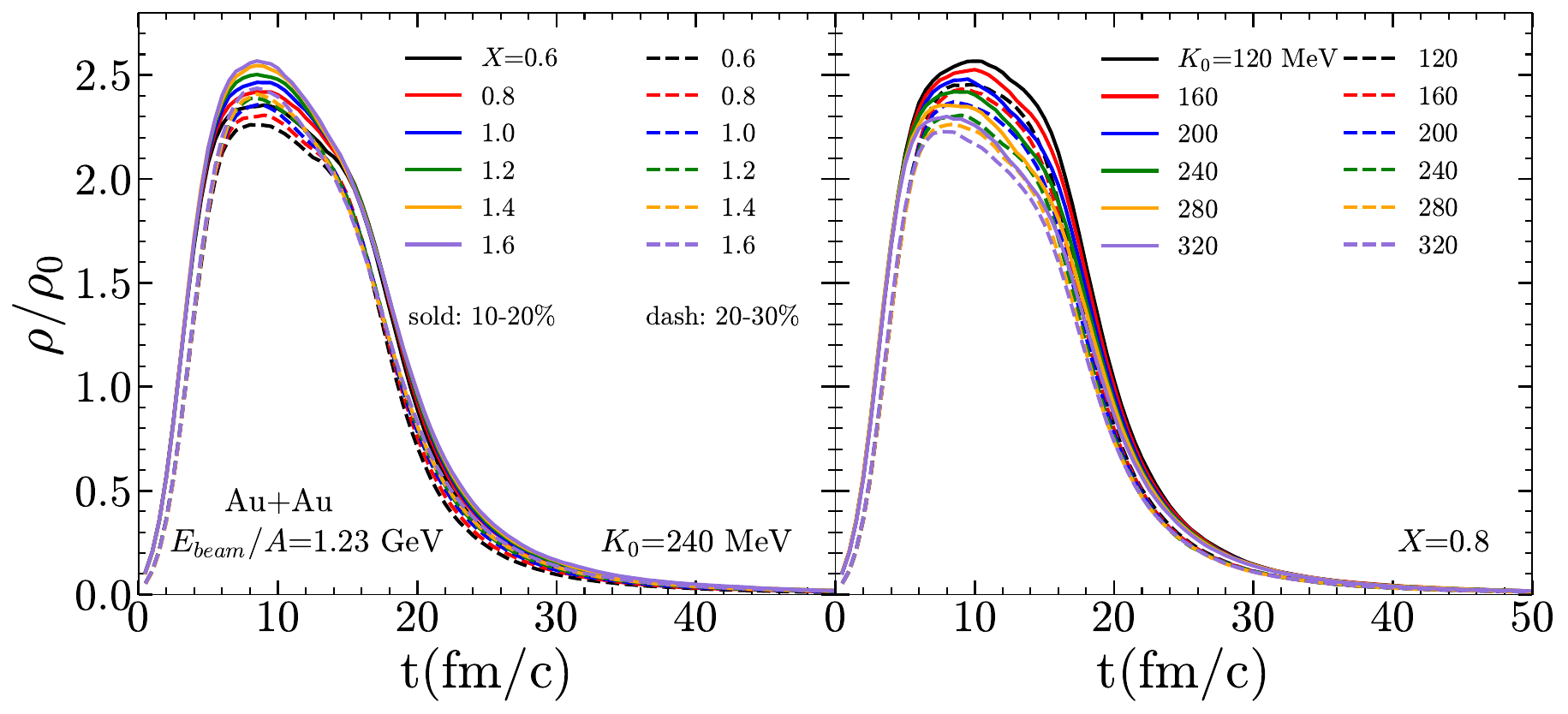}
    \caption{Time evolution of central density in Au + Au collisions at 1.23 GeV for two centrality $10$--$20\%$ and $20$--$30\%$.}
    \label{fig:rho_time}
\end{figure*}

In this section, we first report and discuss the simulation results of Au + Au collisions at $E_{\rm{beam}}/A=1.23~\rm{GeV}$ for two centrality classes, $10$--$20\%$ and $20$--$30\%$, using the IBUU transport model. 
In our simulations, the in-medium cross section modification factor $X$ was varied from 0.5 to 2.0, with a total of 16 discrete values. 
The incompressibility $K_0$ was varied from 100 to 380 MeV, with 15 values in total. 
As a result, for each centrality case, we performed simulations of the Au + Au reactions under 240 different parameter sets of the $X$ and $K_0$, which serve as the training dataset. 
In the Bayesian analysis, however, the values of $X$ and $K_0$ are randomly generated uniformly within their respective prior ranges during the Markov Chain Monte Carlo (MCMC) sampling.  
To investigate how the centrality, in-medium scattering cross section, and incompressibility $K_0$ influence the central density during the reaction, we present in Fig.~\ref{fig:rho_time} the time evolution of the central baryon density for several representative values of $X$ and $K_0$, obtained from the IBUU transport model simulations. 
The solid and dashed lines correspond to the centrality $10$--$20\%$ and $20$--$30\%$, respectively. 
Since collisions at $10$--$20\%$ centrality are more central with a larger number of participating nucleons, the central region of the reaction reaches a higher density.
In the left panel, we fix the incompressibility at $K_0=240$ MeV and compare the effects of different values of the in-medium modification factor $X$ on the central compression density. 
It is observed that the central density increases noticeably with larger values of $X$.
It is because a larger $X$ increases the friction, slowing the interpenetration of the nuclei and allowing for a longer, denser compression phase.
However, the sensitivity of the central density to the medium effect of the cross-section becomes weaker as $X$ increases.
Similarly, the right panel shows the impact of varying $K_0$ values on the central density with $X = 0.8$. 
Because larger $X$ increases the stopping power, while an increase of $K_0$ enhances the repulsive interactions that prevent compression, larger values of $K_0$ correspond to lower central densities. 
This indicates an inverse correlation between the effects of $X$ and $K_0$ on the central density.

\begin{figure*}
    \centering    
    \includegraphics[width=0.8\textwidth]{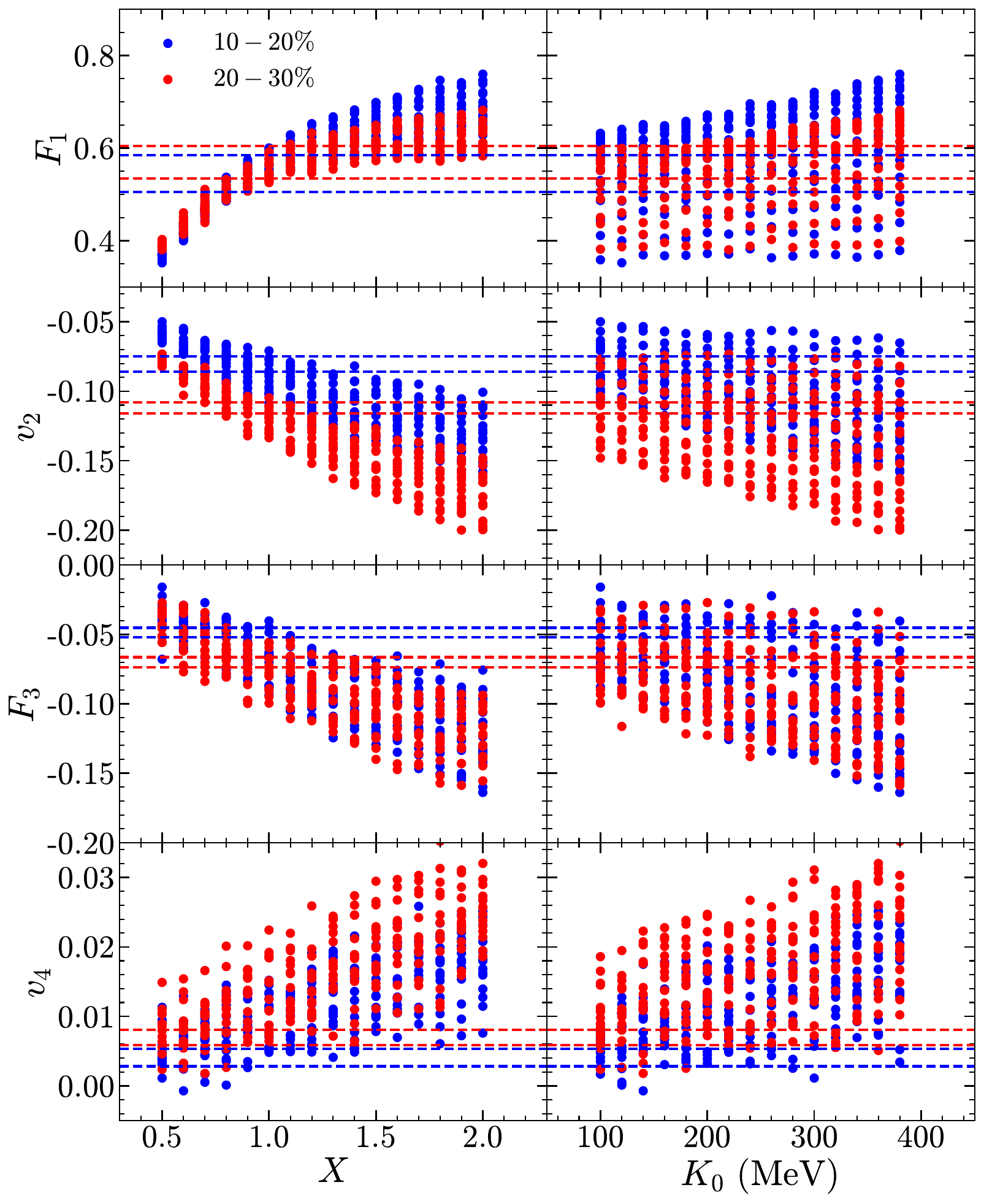}
    \caption{Predicted slope $F_1$ and $F_3$ of the directed flow $v_1$ and triangular flow $v_3$ at mid-rapidity, as well as the elliptic flow $v2$ and quadrupole flow $v_4$ for the free proton, respectively, as functions of the incompressibility $K_0$ and in-medium cross section modification factor $X$ generated by using the IBUU transport model for the Au + Au reactions for two centrality $10$--$20\%$ (blue) and $20$--$30\%$ (red). The dashed line is experimental data from the HADES Collaboration~\cite{2023EPJA...59...80A}.}
    \label{fig:BUU-v1v2}
\end{figure*}

The collective flow of final-state nucleons originates from the fireball region during the reaction and directly reflects the pressure generated in HIC, which is closely related to the density gradient in the reaction. 
In Fig. \ref{fig:rho_time}, we have analyzed the influence of centrality, in-medium cross sections, and the incompressibility of nuclear matter on the central density during the reaction.
In the following, we present the impact of centrality, in-medium cross sections, and the incompressibility of nuclear matter on the collective flow of final-state protons. 
Fig. \ref{fig:BUU-v1v2} displays the slopes $F_1$ and $F_3$ of directed flow and triangular flow of protons at mid-rapidity, as well as the proton elliptic flow $v_2$ and quadrupole flow $v_4$, as functions of the in-medium modification factor $X$ and the incompressibility $K_0$. 
The blue points correspond to the results with centrality $10$--$20\%$, and the red data points represent the $20$--$30\%$ case. 
The blue and red dashed lines indicate the upper and lower bounds of the experimental data from the HADES Collaboration~\cite{2023EPJA...59...80A} for the respective centrality $10$--$20\%$ and $20$--$30\%$.
The left column of Fig. \ref{fig:BUU-v1v2} shows the dependence of $F_1$, $v_2$, $F_3$, and $v_4$ on the in-medium cross-section modification factor $X$, while the right column illustrates their dependence on the incompressibility $K_0$. 
The collective flows $v_1$, $v_2$, $v_3$, and $v_4$ of free protons are functions of rapidity $y$ and transverse momentum $p_t$. 
Specifically, $F_1$ and $F_3$ denote the slopes of $v_1$ and $v_3$ at mid-rapidity $y=0.0$, and $v_2$ and $v_4$ are extracted within the rapidity window $|y_{\rm{cm}}|<0.05$. 
All observables are calculated within the transverse momentum range $0.6\leq p_{\rm{t}}\leq0.9$ GeV/c. 
This interval provides sufficient particle yield to minimize statistical uncertainties in the Bayesian analysis. 

As shown in Fig. \ref{fig:BUU-v1v2}, $F_1$ and $F_3$ are clearly less sensitive to changes in centrality compared to $v_2$ and $v_4$. 
Taking $F_1$ and $v_2$ as examples, for the centrality $20$--$30\%$, the spectators cause stronger shadowing for particles emitted in the $x$-direction than in the $10$--$20\%$ case, which makes the particles more likely to be emitted in the $y$-direction. 
The absolute values of $F_1$, $v_2$, $F_3$, and $v_4$ increase with both the in-medium cross section modification factor $X$ and the incompressibility $K_0$. 
Compared to $K_0$, the collective flows show greater sensitivity to changes in $X$. 
From $F_1$ to $v_4$, the overlap region between the IBUU transport model predictions and the experimental data gradually shifts toward smaller values of $X$.
In the distribution of $F_1$ as a function of $X$, although $F_1$ increases with increasing $X$, the rate of increase slows down. 
This can be attributed to the decreasing growth trend of the maximum central density with increasing $X$, as shown above in Fig. \ref{fig:rho_time}. 
This effect is more pronounced in the centrality $20$--$30\%$ case. 
At lower $X$ values, $F_1$ for centrality $20$--$30\%$ is slightly larger than that for $10$--$20\%$, whereas at higher $X$ values, the opposite trend is observed. 
This can be understood as follows: when $X$ is small, the collision frequency between particles is low, and the contribution to $F_1$ from such collisions is limited. 
The attractive effect of the spectators in the $x$-direction becomes dominant, making $F_1$ slightly larger for centrality $20$--$30\%$. 
Conversely, as $X$ increases, the contribution from the collision between particles becomes more significant, resulting in larger $F_1$ for centrality $10$--$20\%$ than for $20$--$30\%$. 
However, the HADES experimental data show that $F_1$ for centrality $20$--$30\%$ is slightly higher than that for $10$--$20\%$ (the red dashed line is slightly above the blue), suggesting that adopting excessively large $X$ values may contradict the experimental behavior. 
Furthermore, the distribution of $F_1$ becomes broader with increasing $X$, indicating an enhanced sensitivity of $F_1$ to the incompressibility $K_0$. 
A similar trend is also observed in the $F_1$-$K_0$ distribution.

\subsection{Bayesian posterior distributions}

\begin{figure*}
    \centering    
    \includegraphics[width=0.8\textwidth]{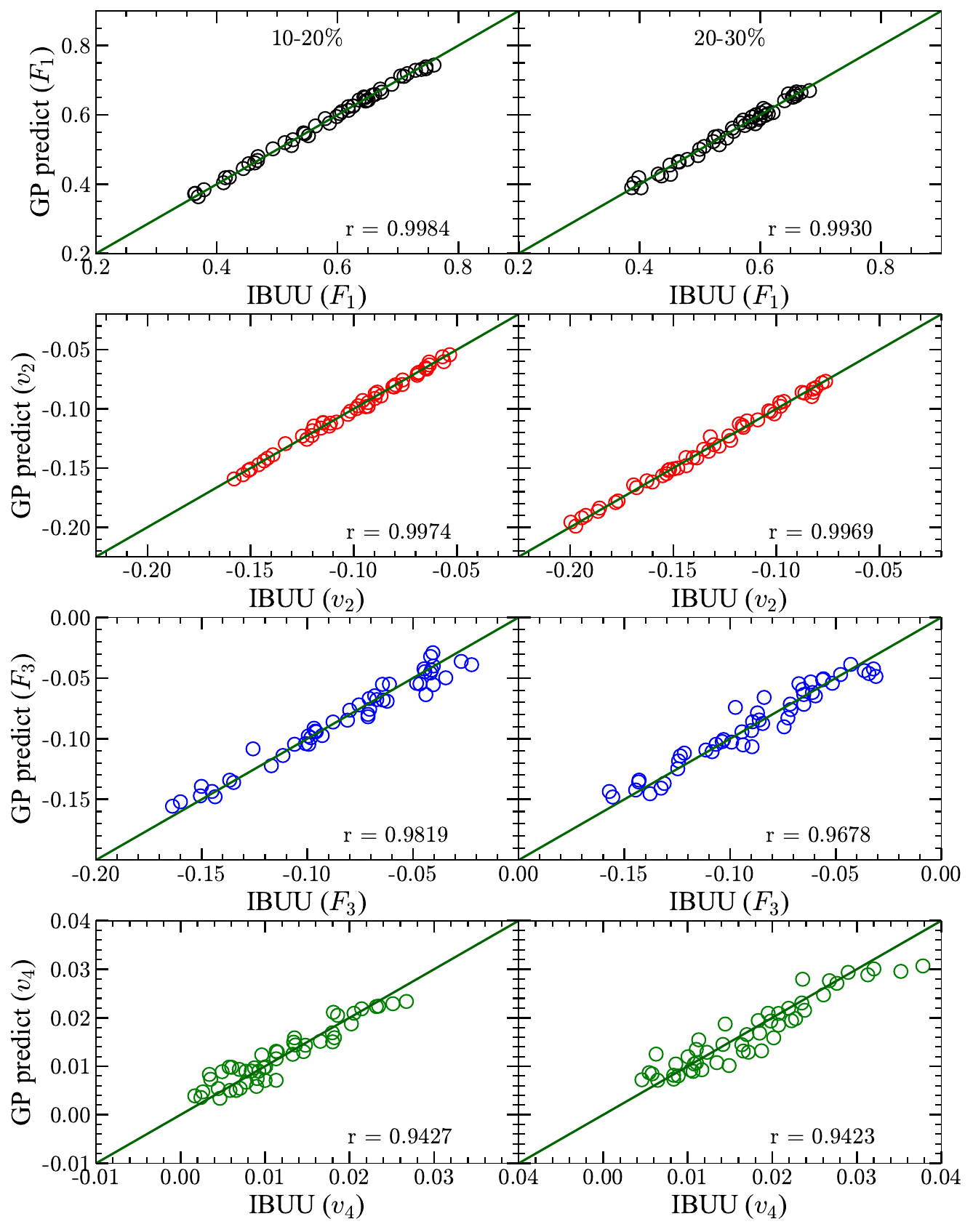}
    \caption{Validation of Gaussian Process (GP) emulations against IBUU predictions for $F_1$, $v_2$, $F_3$, and $v_4$ in mid-central Au+Au collisions. Perfect agreement corresponds to points lying on the $y = x$ line.
    }
    \label{fig:GP}
\end{figure*}

\begin{figure*}
    \centering    
    \includegraphics[width=0.8\textwidth]{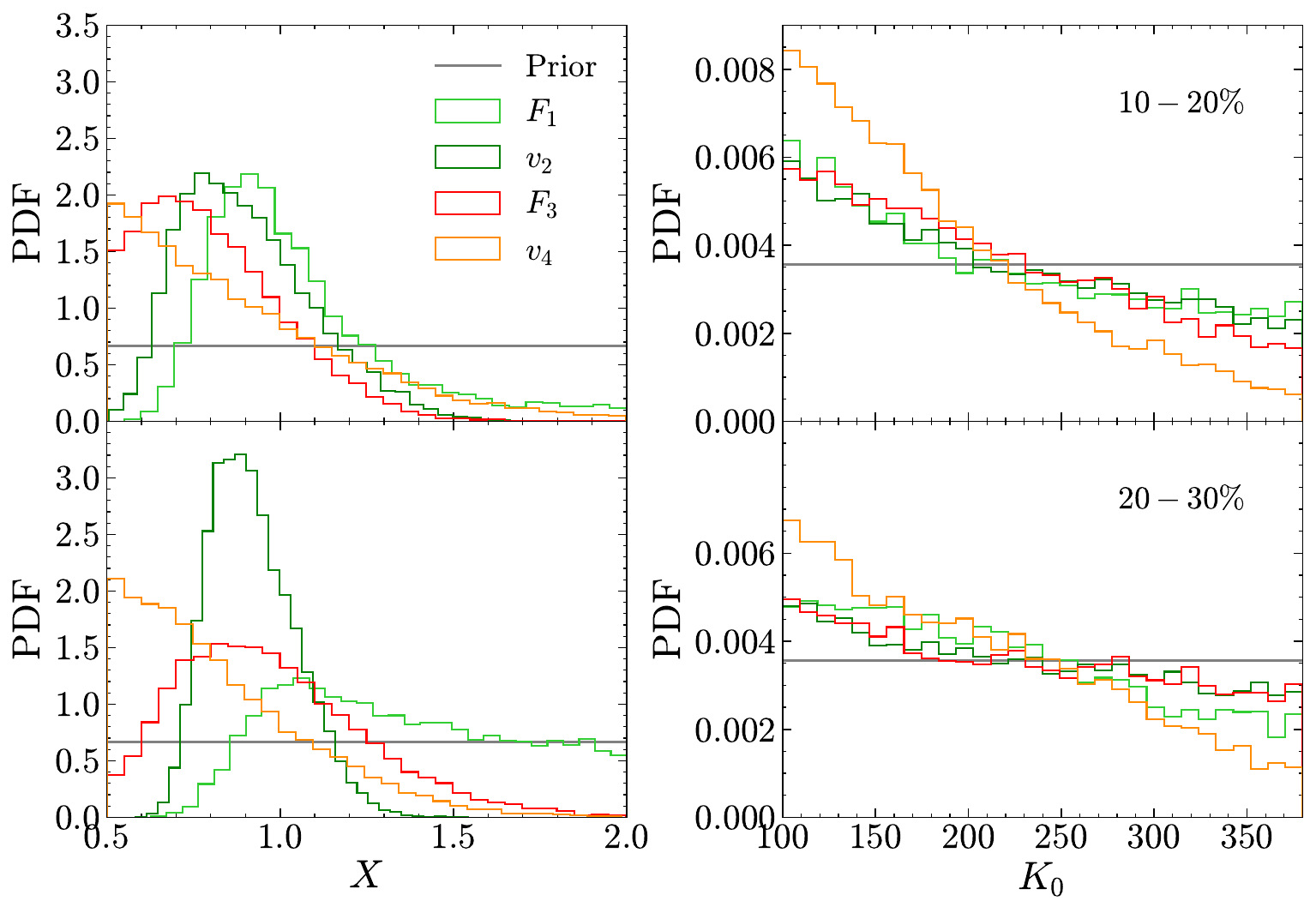}
    \caption{Posterior PDFs of parameters $X$ and $K_0$, from the collective flow observables $F_1$, $v_2$, $F_3$, and $v_4$, for the centrality $10$--$20\%$ (upper panels) and $20$--$30\%$ (lower panels) of Au + Au collisions, respectively.}
    \label{fig:Bay-v1v2v3v4}
\end{figure*}

\begin{table*}[htbp]
\squeezetable
\centering
\caption{Bayesian-inferred posterior results of $X$ and $K_0$ (MeV) from HADES data of proton flow observables $F_1$, $v_2$, $F_3$, and $v_4$ across two centrality intervals: $10$--$20\%$ and $20$--$30\%$.} \label{Tab:Bay-v1v2v3v4}
\setlength{\tabcolsep}{1.8mm}
\renewcommand\arraystretch{1.5}
\begin{ruledtabular}
\begin{tabular}{@{\hspace{5mm}}lcccc@{\hspace{5mm}}}
& \multicolumn{2}{c}{$10$--$20\%$} & \multicolumn{2}{c}{$20$--$30\%$} \\
\cline{2-3} \cline{4-5}
\textbf{Flow} & $X$ & $K_0~(\mathrm{MeV})$ & $X$ & $K_0~(\mathrm{MeV})$ \\
\hline
$F_1$ & $0.98^{+0.30}_{-0.16}$ & $205.42^{+111.95}_{-78.69}$ & $1.23^{+0.49}_{-0.32}$ & $204.68^{+104.51}_{-70.95}$ \\
\hline
$v_2$ & $0.88^{+0.22}_{-0.16}$ & $206.50^{+107.96}_{-77.28}$ & $0.91^{+0.14}_{-0.11}$ & $222.20^{+101.79}_{-87.68}$ \\
\hline
$F_3$ & $0.77^{+0.24}_{-0.17}$ & $199.35^{+101.58}_{-70.91}$ & $0.94^{+0.30}_{-0.23}$ & $222.78^{+101.03}_{-88.70}$ \\
\hline
$v_4$ & $0.81^{+0.40}_{-0.23}$ & $169.99^{+91.22}_{-50.51}$ & $0.77^{+0.32}_{-0.19}$ & $193.72^{+93.94}_{-69.01}$ \\
\end{tabular}
\end{ruledtabular}
\end{table*}

Although we have employed 240 parameter sets of $X$ and $K_0$, they are still insufficient for a Bayesian analysis. 
Given that invoking the momentum-dependent IBUU transport model during the Bayesian inference process is computationally expensive and impractical, we employ the widely used Gaussian Process (GP) emulator. 
To ensure the reliability of the GP emulator, we perform a validation test. 
Specifically, $80\%$ of the dataset is used for training, while the remaining $20\%$ serves as the testing set. 
The kernel function adopted is a combination of the radial basis function (RBF) and white noise kernel.
Figure~\ref{fig:GP} shows the correlation between the predictions from the IBUU transport model and those from the GP emulator for the observables $F_1$, $v_2$, $F_3$, and $v_4$. 
The plots in the left and right columns correspond to two different centrality cases. 
Data points lying along the diagonal line ($y = x$) indicate that the GP emulator can accurately reproduce the results of the IBUU model. 
It is observed that the data points for $F_1$ and $v_2$ lie closely along the $y = x$ lines, indicating high prediction accuracy by the GP emulator. 
While $F_3$ and $v_4$ exhibit a slightly broader spread around the diagonal line. 
This is attributed to the larger model uncertainties associated with higher-order flow observables. 
To further quantify the agreement, the Pearson correlation coefficients (PCC) are also indicated in the Fig. \ref{fig:GP}. 
The PCC between the collective flow predicted by the GP emulator and that generated by the IBUU transport model is defined as
\begin{equation}
r = \frac{\sum_{i=1}^N (x_i - \bar{x})(y_i - \bar{y})}{\sqrt{\sum_{i=1}^N (x_i - \bar{x})^2} \sqrt{\sum_{i=1}^N (y_i - \bar{y})^2}},\label{PCC}
\end{equation}
where $x_i$ and $y_i$ represent the values predicted by the GP emulator and the corresponding values from the IBUU model, respectively. 
$\bar{x}$ and $\bar{y}$ denote their respective means. 
The closer $r$ is to 1.0, the stronger the linear correlation between the two sets of results.
For $F_1$ and $v_2$, the values of $r$ are both greater than 0.99. 
Although $F_3$ and $v_4$ exhibit a small spread, their $r$ values are still above 0.94. 
This further demonstrates that the GP emulator is in strong overall agreement with the results of the IBUU transport model, giving us confidence in the accuracy of our GP emulator.

Figure~\ref{fig:Bay-v1v2v3v4} summarises the Bayesian posteriors PDFs for the in‑medium modification factor $X$ and the incompressibility $K_0$ obtained from the collective flow observables $F_1$, $v_2$, $F_3$, and $v_4$, respectively. 
The upper and lower panels correspond to centralities of $10$--$20\%$ and $20$--$30\%$, while the left and right panels show the posterior distributions of $X$ and $K_0$, respectively. 
The black horizontal lines indicate the prior distributions of $X$ and $K_0$. 
All posteriors deviate markedly from their priors, demonstrating that the flow data provide meaningful constraints on both $X$ and $K_0$.  
When comparing the left and right panels, the posterior distributions of $X$ are more concentrated, suggesting a stronger sensitivity of collective flow to the in-medium correction factor $X$. 
This is similar to the Bayesian analysis results for the medium reduction factor $X$ and the incompressibility $K_0$ using the UrQMD model for FOPI data, as presented in~\cite{2025arXiv250903406W}.
Moreover, the posterior values of $X$ are mostly clustered around 1.0, but slightly below, implying that the scattering cross section tends to be suppressed by in-medium effects. 
This suppression, however, appears relatively weak in Au + Au collisions at 1.23~GeV/nucleon, which is consistent with previous findings from BHF calculations~\cite{2022PhRvC.106f4332H} and the UrQMD transport model analysis~\cite{2022PhLB..82837019L} of FOPI data, where the in-medium correction factor approaches 1.0 at higher energies.
From $F_1$ to $v_4$, the $X$ posteriors gradually shift toward smaller values, consistent with the trend seen in Fig.~\ref{fig:BUU-v1v2}. 
In the centrality $20$--$30\%$, due to the influence of spectator matter, the observable $F_1$ still allows a considerable portion of the $X$ distribution in the region above 1.0. 
In contrast, $v_2$ provides a much tighter constraint on $X$, reflecting its stronger sensitivity to in-medium modifications. 
Although the posterior distributions of $K_0$ are relatively broad, there is still a noticeable accumulation below about $250~\rm{MeV}$, implying that collective flow data can help distinguish extremely stiff or soft EoS, with a tendency toward softer EoS. 
The posteriors of $K_0$ are only weakly dependent on centrality, with minor differences observed between the two centralities. 
Among the four observables, the constraint on $K_0$ derived from $v_4$ shows a notable deviation compared to those from the other observables. 
This may be attributed to the larger model uncertainty in predicting $v_4$ within the IBUU transport model, as well as the significant experimental uncertainty in the measurement of $v_4$. 
For both centralities, Table~\ref{Tab:Bay-v1v2v3v4} summarizes the values of $X$ and $K_0$ within the $1\sigma$ credible intervals constrained by $F_1$, $v_2$, $F_3$, and $v_4$, respectively.

\begin{figure*}
    \centering    
    \includegraphics[width=0.48\textwidth]{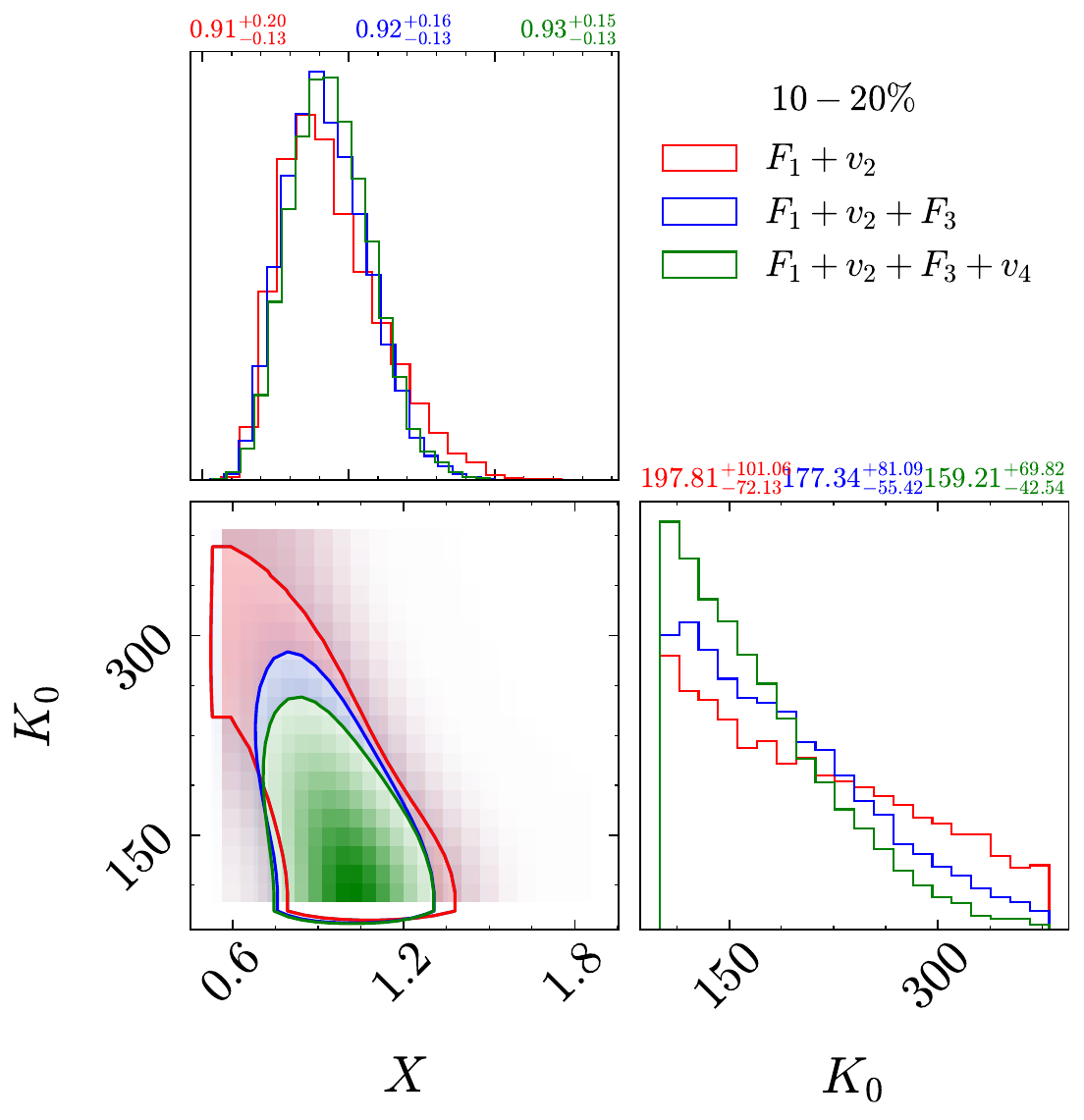}
    \includegraphics[width=0.48\textwidth]{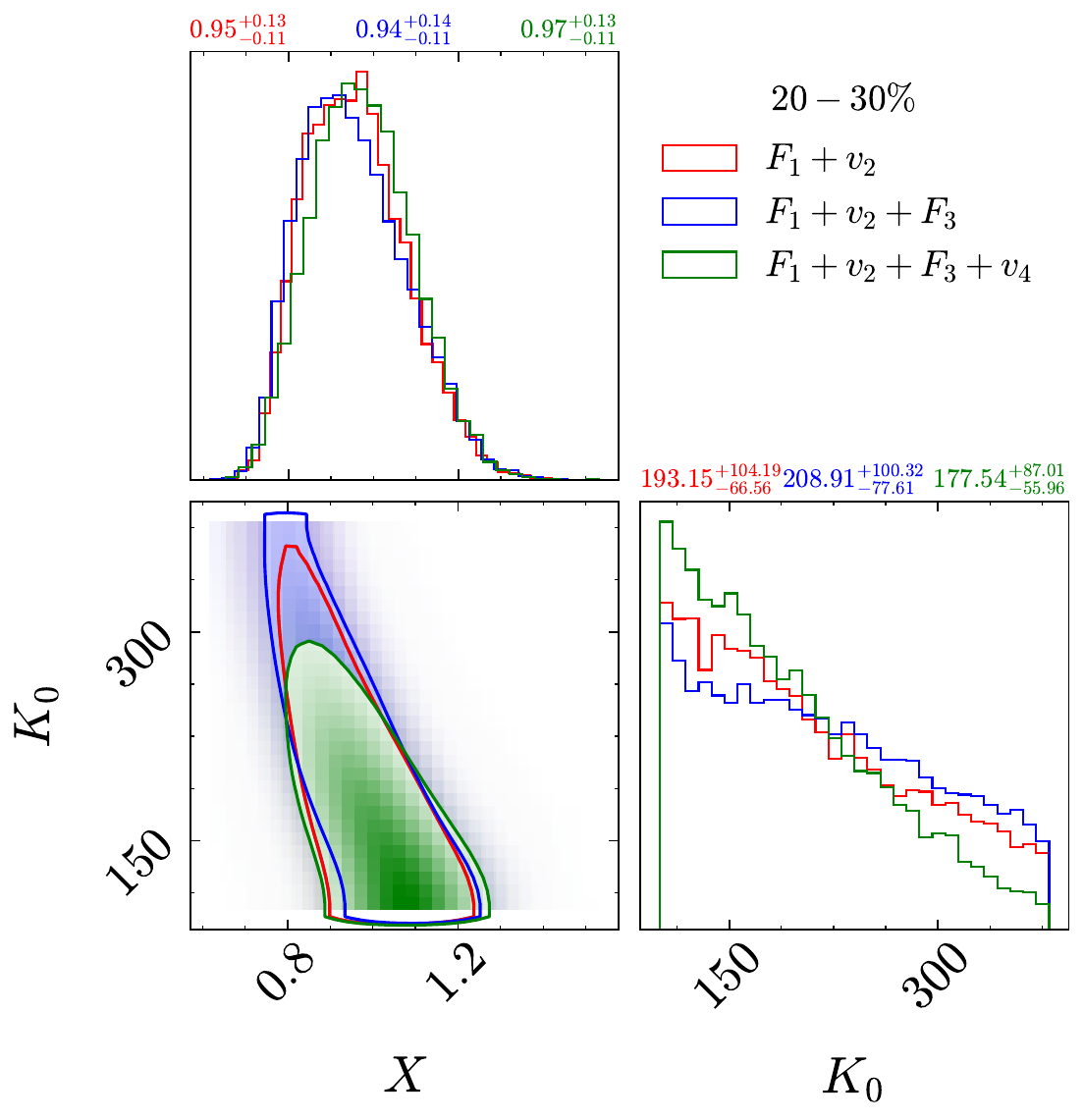}
    \caption{Posterior PDFs of $X$ and $K_0$ as well as their correlation, for the centrality $10$--$20\%$ (left panels) and $20$--$30\%$ (right panels) of Au + Au collisions, respectively.}
    \label{fig:Bay_2D}
\end{figure*}

\begin{table*}[htbp]
\squeezetable
\centering
\caption{Posterior results of the in-medium baryon-baryon cross-section factor $X$ and nuclear incompressibility $K_0$ (MeV), extracted via Bayesian analysis of HADES proton flow data. Results are shown for three combinations of observables ($F_1+v_2$, $F_1+v_2+F_3$, and $F_1+v_2+F_3+v_4$) across two centrality intervals (10–20\% and 20–30\%) considering the MDI (momentum-dependent interaction). For comparison, posterior values of $X$ and $K_0$ obtained using a momentum-independent mean field are also presented.
} \label{mean values}
\setlength{\tabcolsep}{1.8mm}
\renewcommand\arraystretch{1.5}
\begin{ruledtabular}
\begin{tabular}{lcccc}
 & \multicolumn{2}{c}{10--20\%} & \multicolumn{2}{c}{20--30\%} \\
\cline{2-3} \cline{4-5}
\textbf{Flow combination} & $X$ & $K_0$ (MeV) & $X$ & $K_0$ (MeV) \\
\hline
\multicolumn{5}{c}{\textbf{With MDI (momentum-dependent interaction)}} \\
$F_1+v_2$                & $0.91^{+0.20}_{-0.13}$ & $197.8^{+101.1}_{-72.1}$ & $0.95^{+0.13}_{-0.11}$ & $193.2^{+104.2}_{-66.6}$ \\
$F_1+v_2+F_3$            & $0.92^{+0.16}_{-0.13}$ & $177.3^{+81.1}_{-55.4}$  & $0.94^{+0.14}_{-0.11}$ & $208.9^{+100.3}_{-77.6}$ \\
$F_1+v_2+F_3+v_4$        & $0.93^{+0.15}_{-0.13}$ & $159.2^{+69.8}_{-42.5}$  & $0.97^{+0.13}_{-0.11}$ & $177.5^{+87.0}_{-56.0}$ \\
\hline
\multicolumn{5}{c}{\textbf{Without MDI (momentum-independent mean field)}} \\
$F_1+v_2$                & $1.36^{+0.41}_{-0.38}$ & $204.2^{+97.6}_{-45.0}$  & $1.55^{+0.28}_{-0.32}$ & $252.1^{+68.2}_{-36.7}$ \\
$F_1+v_2+F_3$            & $1.42^{+0.37}_{-0.42}$ & $188.6^{+86.9}_{-36.0}$  & $1.59^{+0.27}_{-0.34}$ & $238.8^{+62.4}_{-30.6}$ \\
$F_1+v_2+F_3+v_4$        & $1.33^{+0.38}_{-0.33}$ & $195.2^{+77.6}_{-41.1}$  & $1.57^{+0.27}_{-0.34}$ & $237.3^{+63.0}_{-31.3}$ \\
\end{tabular}
\end{ruledtabular}
\end{table*}

Given the evident differences in how $F_1$, $v_2$, $F_3$, and $v_4$ constrain the in-medium cross section, as well as the considerable uncertainties in both model predictions and experimental measurements for higher-order collective flows, we further consider three combined cases for constraining $X$ and $K_0$: (1) $F_1 + v_2$, (2) $F_1 + v_2 + F_3$, and (3) $F_1 + v_2 + F_3 + v_4$. 
Figure~\ref{fig:Bay_2D} presents the two-dimensional posterior distributions of $X$ and $K_0$ under these three constraint cases.
The left panel corresponds to the centrality range of $10$--$20\%$, while the right panel corresponds to $20$--$30\%$. 
The $X$-$K_0$ correlation plots exhibit a noticeable anti-correlation between the two parameters. 
This anti-correlation emerges because both a stiffer EoS (higher $K_0$) and a larger cross section (higher $X$) increase the nuclear stopping power and pressure buildup, leading to stronger collective flow. Therefore, the data can be fit equally well by a combination of a soft EoS with a large cross section or a stiff EoS with a small cross section. Our Bayesian analysis breaks this degeneracy by identifying the most probable region within this valley of correlation, which points towards the soft EoS/mild suppression scenario.
It should be noted that the IBUU transport model employs test particles, which suppress the effects of fluctuations and therefore impact higher-order flows and flows in more central collisions.
Since the constraining power weakens progressively from $F_1$ to $v_4$ (see Fig.~\ref{fig:Bay-v1v2v3v4}), the posterior values of $X$ under the three combined cases all lie between those determined by $F_1$ and $v_2$ individually.
For the centrality $20$--$30\%$ case, where 
$v_2$ offers significantly stronger constraints on $X$ than the other observables, and the posterior distribution becomes more concentrated. 
The posterior values of $X$ and $K_0$ under the three combined constraint cases for both centralities are listed in Table~\ref{mean values}. 
The anti-correlation between $X$ and $K_0$ can also be inferred from their values. 
In all three cases, it is found that the values of $X$ are quite close and slightly less than 1.0, while the values of $K_0$ are all below 210 MeV, indicating a soft EoS. 
In Table~\ref{mean values}, we also present the Bayesian-inferred values of $X$ and $K_0$ obtained with a momentum-independent mean field, which are higher than those with a momentum-dependent mean field.
This is consistent with the findings of Ref.~\cite{2023NuPhA103922726L}, where a much larger medium correction factor, $X = 1.32^{+0.28}_{-0.40}$, and a stiffer EoS with $K_0 = 346^{+29}_{-31}$ MeV were obtained using a momentum-independent mean field.
The discrepancy arises because, within transport models, a larger incompressibility $K_0$ is generally required when using a momentum-independent potential in order to reproduce similar observables as those obtained with a momentum-dependent potential~\cite{1987PhRvC..35.1666G,1988PhRvC..38.2101W,2023NuPhA103922726L,2024arXiv240607051W}.
Moreover, for the centrality $20$--$30\%$ case, the posterior values of $X$ tend to be higher than those for $10$--$20\%$, which can be attributed to the influence of spectator matter on the constraint from $F_1$. 

\section{Summary}\label{Sec:Summary}

In summary, we have employed a Bayesian statistical framework, combined with Gaussian Process emulators and the IBUU transport model incorporating a momentum-dependent single-particle potential, to extract constraints on the nuclear incompressibility $K_0$ and the in-medium baryon-baryon cross-section modification factor $X$. The analysis is based on collective flow data of free protons measured in 1.23 GeV/nucleon Au+Au collisions by the HADES Collaboration.

We considered two centrality classes ($10$–$20\%$ and $20$–$30\%$), using free proton observables that include higher-order flow components—namely, the slopes of directed and triangular flow ($F_1$, $F_3$), as well as elliptic and quadrupole flow ($v_2$, $v_4$). In the parameter space defined by $X \in [0.5, 2.0]$ and $K_0 \in [100, 380]$ MeV, 
we trained the GP emulator with 240 parameter sets evaluated within the IBUU model. 
 
Our results show that collective flow observables are more sensitive to the variation in $X$, with HADES data generally favoring $X < 1.3$. 
Moreover, higher-order flows prefer smaller values of $X$, with Bayesian posteriors indicating a most probable range of $X \sim 0.9$–$1.0$, suggesting mild in-medium suppression of $NN$ cross sections. 
Our results, favoring $X \approx 1.0 $, suggest that the in-medium suppression at 1.23 GeV might be weaker than predicted by some zero-temperature microscopic calculations. This could be due to energy-dependent effects, the inclusion of inelastic channels, or a compensation mechanism between different medium-modification factors not captured by our simple scaling.
At the same time, the data favor a soft to moderately soft EoS, with $K_0$ most likely below 250 MeV, but do not strongly rule out stiffer values within the $1\sigma$ credible interval.
Comparative analysis across the two centralities reveals that spectator matter can influence the posterior distribution of $X$, particularly in more peripheral collisions. 
Also, compared with the results obtained using a momentum-independent mean field, we find that the Bayesian-inferred values of $X$ and $K_0$ are smaller when a momentum-dependent mean field is employed.
This is because the momentum dependence of the mean field already enhances the nuclear stopping power in HICs, thereby reducing the need for large values of $X$ and $K_0$ to reproduce the experimental data. This direct comparison demonstrates that momentum dependence in the mean field naturally enhances stopping and collective flow, thereby reducing the need to invoke a stiff EoS or excessively large cross-section scaling. 

For continued progress toward a consistent extraction of the nuclear mean field and the in-medium $NN$ cross section from HIC data, we plan to move beyond simple scalings. This involves implementing density- and momentum-dependent parameterizations informed by microscopic calculations, and explicitly incorporating both model and emulator uncertainties into the Bayesian posterior inference to obtain more reliable parameter constraints. It is equally important to expand the set of observables to include isospin-sensitive probes, such as neutron-to-proton flow ratios and $\pi^-/\pi^+$ yield ratios, which exhibit different sensitivities to the incompressibility $K_0$ and symmetry-energy parameters. In addition, extending the analysis to collision systems across a broad beam-energy range (0.1--3 GeV/nucleon) will be essential for constraining both the low- and high-density behavior of the nuclear EoS and in-medium $NN$ interactions.

\section*{Acknowledgments}
We gratefully acknowledge Bao-An Li for his constructive suggestions and for carefully reviewing the manuscript. The work is supported by the National Natural Science Foundation of China (grant Nos.~12273028, 12494572). 

\section*{Data Availability}
The data that support the findings of this article are openly available~\cite{2023EPJA...59...80A,2017NuPhA.967..812K,2020PhRvL.125z2301A}. 

\appendix 
\section{} \label{appendixA}

For the ImMDI based on the Hartree–Fock calculation using the Gogny interaction, the baryon potential energy density can be expressed as
\begin{align}
    V_{\rm ImMDI}(\rho,\beta)&=\frac{A_u\rho_n\rho_p}{\rho_0}+\frac{A_l}{2\rho_0}(\rho_n^2+\rho_p^2)\nonumber \\
    &+\frac{B}{\sigma+1}\frac{\rho^{\sigma+1}}{\rho_0^\sigma}(1-x\beta^2)\nonumber \\  &+\sum_{\tau,\tau'}\frac{C_{\tau,\tau'}}{\rho_0}\int\int d^3pd^3p'\frac{f_{\tau}(\vec{r},\vec{p})f_{\tau'}(\vec{r},\vec{p}')}{1+(\vec{p}-\vec{p}')^2/\Lambda^2}.
\end{align}
At zero temperature, the phase space distribution function can be expressed using the step function as $f_{\tau}(\vec{r},\vec{p})=\frac{2}{h^3}\Theta(p_f(\tau)-p)$. The integral expression in the momentum-dependent term can then be calculated analytically as \cite{2017PhRvC..95c4324K}    
\begin{align}
    &\int d^3p'\frac{f_{\tau}(\vec{r},\vec{p}')}{1+(\vec{p}-\vec{p}')^2/\Lambda^2}\nonumber \\
    &=\frac{2\pi\Lambda^3}{h^3}\Big\{ \frac{p_{f\tau}^2+\Lambda^2-p^2}{2p\Lambda}\ln{\Big[ \frac{(p+p_{f\tau})^2+\Lambda^2}{(p-p_{f\tau})^2+\Lambda^2} \Big]}\nonumber \\
    &+\frac{2p_{f\tau}}{\Lambda}-2\Big( \arctan{\frac{p+p_{f\tau}}{\Lambda}}-\arctan{\frac{p-p_{f\tau}}{\Lambda}} \Big) \Big\},
\end{align}
and
\begin{align}  
    &\int\int d^3pd^3p'\frac{f_{\tau}(\vec{r},\vec{p})f_{\tau'}(\vec{r},\vec{p}')}{1+(\vec{p}-\vec{p}')^2/\Lambda^2}\nonumber \\
    &=\frac{1}{6}\Big( \frac{4\pi}{h^3} \Big)^2\Lambda^2\Big\{ p_{f\tau}p_{f\tau'}\big[ 3(p_{f\tau}^2+p_{f\tau'}^2)-\Lambda^2  \big]\nonumber \\
    &+4\Lambda\Big[ (p_{f\tau}^3-p_{f\tau'}^3)\arctan{\Big( \frac{p_{f\tau}-p_{f\tau'}}{\Lambda} \Big)}\nonumber \\ 
    &-(p_{f\tau}^3+p_{f\tau'}^3)\arctan{\Big( \frac{p_{f\tau}+p_{f\tau'}}{\Lambda} \Big)} \Big]\nonumber \\
    &+\frac{1}{4}\big[ \Lambda^4+6\Lambda^2(p_{f\tau}^2+p_{f\tau'}^2)-3(p_{f\tau}^2-p_{f\tau'}^2)^2 \big] \nonumber \\
    &\times\ln{\Big[ \frac{(p_{f\tau}+p_{f\tau'})^2+\Lambda^2}{(p_{f\tau}-p_{f\tau'})^2+\Lambda^2} \Big]} \Big\}.
\end{align} 

The EoS of cold symmetric nuclear matter \cite{2017PhRvC..95c4324K}:
\begin{align}
    E_0(\rho)&=\frac{8\pi}{5mh^3\rho}p_f^5+\frac{\rho}{4\rho_0}(A_{l}+A_{u})+\frac{B}{\sigma+1}\Big( \frac{\rho}{\rho_0} \Big)^{\sigma}\nonumber \\
    &+\frac{C_{l}+C_{u}}{3\rho_0\rho}\Big( \frac{4\pi}{h^3} \Big)^2\Lambda^2\Big[ p_f^2(6p_f^2-\Lambda^2)\nonumber \\
    &-8\Lambda p_f^3 \arctan{\Big( \frac{2p_f}{\Lambda} \Big)}\nonumber \\
    &+\frac{\Lambda^4+12\Lambda^2 p_f^2}{4}\ln{\Big( \frac{4p_f^2+\Lambda^2}{\Lambda^2} \Big)} \Big].\label{2}
\end{align}
The first-order derivative of the binding energy of each nucleon with respect to the density is expressed as \cite{2017PhRvC..95c4324K}
\begin{align}
    \frac{dE_0(\rho)}{d\rho}&=\frac{16\pi p_f^5}{15mh^3\rho^2}+\frac{A_{l}+A_{u}}{4\rho_0}+\frac{B\sigma}{\sigma+1}\frac{\rho^{\sigma-1}}{\rho_0^{\sigma}}\nonumber \\
    &+\frac{C_{l}+C_{u}}{3\rho_0\rho^2}\Big( \frac{4\pi}{h^3} \Big)^2\Lambda^2
    \Big[ 2p_f^4+\Lambda^2p_f^2\nonumber \\
    &-\frac{\Lambda^4+4\Lambda^2p_f^2}{4}\ln{\Big( \frac{4p_f^2+\Lambda^2}{\Lambda^2} \Big)} \Big], \label{3}
\end{align}
and the second-order derivative is expressed as 
\begin{align}
    \frac{d^2E_0(\rho)}{d\rho^2}&=\frac{-16\pi p_f^5}{45mh^3\rho^3}+\frac{B\sigma(\sigma-1)}{\sigma+1}\frac{\rho^{\sigma-2}}{\rho_0^{\sigma}}\nonumber \\
    &+\frac{2(C_{l}+C_{u})}{3\rho_0\rho^3}\Big( \frac{4\pi}{h^3} \Big)^2\Lambda^2\Big[ -\frac{2p_f^4}{3}\nonumber \\
    &-\Lambda^2p_f^2
    +\Lambda^2\Big( \frac{\Lambda^2}{4}+\frac{2p_f^2}{3} \Big)\ln{\Big( \frac{4p_f^2+\Lambda^2}{\Lambda^2} \Big)} \Big]. \label{4}
\end{align}
The incompressibility of symmetric nuclear matter is defined as \cite{2017PhRvC..95c4324K}
\begin{align}
    K_0=9\rho_0^2\Big( \frac{d^2E_0(\rho)}{d\rho^2} \Big)_{\rho=\rho_0}.\label{5}
\end{align}
The symmetry energy of nuclear matter is defined as the second derivative of the binding energy per nucleon with respect to isospin asymmetry, given by \cite{2017PhRvC..95c4324K}
\begin{align}
    E_{\rm sym}(\rho)&=\frac{1}{2}\Big( \frac{\partial^2E(\rho,\beta)}{\partial\beta^2} \Big)_{\beta=0}\nonumber \\
    &=\frac{8\pi p_f^5}{9mh^3\rho}+\frac{\rho(A_{l}-A_{u})}{4\rho_0}-\frac{Bx}{\sigma+1}\Big( \frac{\rho}{\rho_0} \Big)^{\sigma}\nonumber \\
    &+\frac{C_{l}}{9\rho_0\rho}\Big( \frac{4\pi}{h^3} \Big)^2\Lambda^2\Big[ 4p_f^4-\Lambda^2p_f^2\ln{\Big( \frac{4p_f^2+\Lambda^2}{\Lambda^2} \Big)} \Big]\nonumber \\
    &+\frac{C_{u}}{9\rho_0\rho}\Big( \frac{4\pi}{h^3} \Big)^2\Lambda^2\Big[ 4p_f^4\nonumber \\
    &-p_f^2(4p_f^2+\Lambda^2)\ln{\Big( \frac{4p_f^2+\Lambda^2}{\Lambda^2} \Big)} \Big]. \label{6}
\end{align}
The first-order derivative of the symmetry energy with respect to density is expressed as
\begin{align}
    \frac{dE_{\rm sym}(\rho)}{d\rho}&=\frac{16\pi p_f^5}{27mh^3\rho^2}+\frac{A_{l}-A_{u}}{4\rho_0}-\frac{Bx\sigma}{\sigma+1}\frac{\rho^{\sigma-1}}{\rho_0^{\sigma}}\nonumber \\
    &+\frac{C_{l}+C_{u}}{27\rho_0\rho^2}\Big( \frac{4\pi}{h^3} \Big)^2\Lambda^2\Big[ 4p_f^4\nonumber \\
    &+\Lambda^2p_f^2\ln{\Big( \frac{4p_f^2+\Lambda^2}{\Lambda^2} \Big)}-\frac{8\Lambda^2p_f^4}{4p_f^2+\Lambda^2} \Big]\nonumber \\
    &-\frac{4C_u}{27\rho_0\rho^2}\Big( \frac{4\pi}{h^3} \Big)^2\Lambda^2 p_f^4\Big[ \ln{\Big( \frac{4p_f^2+\Lambda^2}{\Lambda^2} \Big)}\nonumber \\
    &+\frac{8p_f^2}{4p_f^2+\Lambda^2} \Big],\label{Eq:dEsym_drho}
\end{align}
and the slope of the symmetry energy at saturation density $\rho_0$ is defined as 
\begin{align}
    L=3\rho_0\Big( \frac{dE_{\rm sym}(\rho)}{d\rho} \Big)_{\rho=\rho_0}.\label{Eq:L0}
\end{align}

In non-relativistic approaches, the effective mass $m^*_{\tau}$ of a nucleon can be defined via the following expressions
\begin{align}
    \frac{m^*_{\tau}}{m_{\tau}}=\Big( 1+\frac{m_{\tau}}{p}\frac{dU_{\tau}(\rho,\beta,\vec{p})}{dp} \Big)^{-1}.\label{7}
\end{align}
At zero temperature, the derivative of the mean field with $\beta=0$ with respect to momentum is expressed as
\begin{align}
    \frac{dU_0(\rho,p)}{dp}&=\frac{2(C_{l}+C_{u})}{\rho_0}\frac{2\pi\Lambda^3}{h^3}\nonumber \\
    &\times\Big[ -(\frac{p_f^2+\Lambda^2+p^2}{2\Lambda p^2})\ln{\Big[ \frac{(p+p_{f})^2+\Lambda^2}{(p-p_{f})^2+\Lambda^2} \Big]} \nonumber \\
    &+\frac{(p_f^2+\Lambda^2-p^2)[4\Lambda^2 p+4p_f(p^2-p_f^2)]}
    {2\Lambda p[(p+p_f)^2+\Lambda^2][(p-p_f)^2+\Lambda^2]}\nonumber \\
    &-\frac{2}{\Lambda}\Big( \frac{\Lambda^2}{\Lambda^2+(p+p_f^2)^2}-\frac{\Lambda^2}{\Lambda^2+(p-p_f^2)^2} \Big) \Big].\label{8}
\end{align}

We choose the following empirical values for the saturation point of symmetric nuclear matter \cite{2015PhRvC..91a4611X, 2017PhRvC..95c4324K}, i.e., the binding energy $E_0=-16$ MeV, the saturation density $\rho_0=0.16$ fm$^{-3}$, the symmetry energy $E_{\rm sym}(\rho_0)=32.5$ MeV, the single-particle potential $U(\rho=\rho_0,\beta=0,p\to\infty)=75$ MeV, the effective mass $m^*/m=0.7$. 
From Eq. (\ref{1}), (\ref{2}), (\ref{3}), (\ref{4}), (\ref{5}), (\ref{6}), (\ref{7}), and (\ref{8}) the following system of equations can be obtained
\begin{widetext}
\begin{equation}
    \begin{cases}
    &U^{\infty}_0=A_{l0}+B=75\nonumber \\
    
    &\Big( \frac{m^*}{m} \Big)_{p=p_{f_0}}=
    1-0.7\frac{2m(C_{l0}+C_{u0})}{\rho_0}\frac{\Lambda_0^3 p_{f_0}}{4\pi^2}\Big[ \frac{2}{\Lambda_0}-(\frac{\Lambda_0}{2}+\frac{1}{\Lambda_0})\ln{\frac{4+\Lambda_0^2}{\Lambda_0^2}} \Big]=0.7\nonumber \\
    
    &E_0(\rho_0)= \frac{p_{f_0}^5}{5m\pi^2\rho_0}+\frac{A_{l0}}{2}+\frac{B}{\sigma+1}+\frac{3(C_{l0}+C_{u0})\Lambda_0^2}{16}\Big[ 6-\Lambda_0^2-8\Lambda_0\arctan{\frac{2}{\Lambda_0}}+\frac{\Lambda_0^4+12\Lambda_0^2}{4}\ln{\frac{4+\Lambda_0^2}{\Lambda_0^2}} \Big]=-16 \nonumber \\
    
    &\Big(\frac{dE_0(\rho_0)}{d\rho}\Big)_{\rho=\rho_0}=\frac{p_{f_0}^2}{5m\rho_0}+\frac{A_{l0}}{2\rho_0}+\frac{B\sigma}{\rho_0(\sigma+1)}+\frac{3(C_{l0}+C_{u0})\Lambda_0^2}{16\rho_0}\Big[2+\Lambda_0^2-\frac{\Lambda_0^4+4\Lambda_0^2}{4}ln\frac{4+\Lambda_0^2}{\Lambda_0^2} \Big]=0
    \nonumber \\
    
    &K_0=-\frac{3p_{f_0}^2}{5m}+\frac{9B(\sigma-1)\sigma}{\sigma+1}+\frac{27\Lambda_0^2(C_{l0}+C_{u0})}{8}\Big[-\frac{2}{3}-\Lambda_0^2+\Lambda_0^2(\frac{\Lambda_0^2}{4}+\frac{2}{3})ln\frac{4+\Lambda_0^2}{\Lambda_0^2} \Big]\nonumber \\
    
    &E_{\rm sym}(\rho_0)=\frac{p_{f_0}^5}{9m\pi^2\rho_0}+\frac{C_{l0}\Lambda^2_0}{16}\Big[4-\Lambda_0^2 ln\frac{4+\Lambda_0^2}{\Lambda_0^2} \Big]+\frac{C_{u0}\Lambda^2_0}{16}\Big[4-(4+\Lambda_0^2) ln\frac{4+\Lambda_0^2}{\Lambda_0^2} \Big]=32.5,  \nonumber \\
    \end{cases}
\end{equation}
\end{widetext}
where $\Lambda_0 p_{f_0}$ is used instead of $\Lambda$. 
By solving the above system of equations, we can get the values of parameters $A_{l0},~ B,~ \sigma,~ C_{l0},~ C_{u0}$ and $\Lambda_0$. Here we use $vpasolve$ in $matlab$ to solve the system of equations.

\bibliography{ref}
\bibliographystyle{apsrev4-1}

\end{document}